# Beyond performance metrics: Examining a decrease in students' physics self-efficacy through a social networks lens

Remy Dou,[1] Eric Brewe,[1,2,3] Justyna P. Zwolak,[1,3] Geoff Potvin,[2]
Eric A. Williams,[2] and Laird H. Kramer[2,3]

[1]*Department of Teaching and Learning, Florida International University,*
*11200 S.W. 8th Street, Miami, Florida 33199, USA*
[2]*Department of Physics, Florida International University,*
*11200 S.W. 8th Sreet, Miami, Florida 33199, USA*
[3]*STEM Transformation Institute, Florida International University,*
*11200 S.W. 8th Street, Miami, Florida 33199, USA*

The Modeling Instruction (MI) approach to introductory physics manifests significant increases in student conceptual understanding and attitudes toward physics. In light of these findings, we investigated changes in student self-efficacy while considering the construct's contribution to the career-decision making process. Students in the Fall 2014 and 2015 MI courses at Florida International University exhibited a decrease on each of the sources of self-efficacy and overall self-efficacy ($N = 147$) as measured by the Sources of Self-Efficacy in Science Courses-Physics (SOSESC-P) survey. This held true regardless of student gender or ethnic group. Given the highly interactive nature of the MI course and the drops observed on the SOSESC-P, we chose to further explore students' changes in self-efficacy as a function of three centrality measures (i.e., relational positions in the classroom social network): inDegree, outDegree, and PageRank. We collected social network data by periodically asking students to list the names of peers with whom they had meaningful interactions. While controlling for PRE scores on the SOSESC-P, bootstrapped linear regressions revealed post-self-efficacy scores to be predicted by PageRank centrality. When disaggregated by the sources of self-efficacy, PageRank centrality was shown to be directly related to students' sense of mastery experiences. InDegree was associated with verbal persuasion experiences, and outDegree with both verbal persuasion and vicarious learning experiences. We posit that analysis of social networks in active learning classrooms helps to reveal nuances in self-efficacy development.

## I. INTRODUCTION

The implementation of active learning environments across science, technology, engineering, and mathematics (STEM) fields has garnered attention from education researchers across the country. Their work has revealed with strong significance the advantage of active learning strategies over traditional, lecture-based pedagogies [1]. Specifically in the arena of physics education, a variety of active learning approaches have led to the reformation of introductory physics courses in colleges and universities. These include Investigative Science Learning Environments (ISLE), Student-Centered Activities for Large Enrollment University Physics (SCALE-UP), Workshop Physics, Tutorials in Introductory Physics, and Modeling Instruction (MI) among others. To various degrees, they have exhibited positive impacts on student learning [2–6].

Yet, success in physics education, particularly in the realm of career persistence, involves more than just improving learning gains; it requires exploring changes in affective constructs that complement academic performance [7–9].

Affective constructs, like self-efficacy, identity, and interest, are often positively correlated with student outcomes like academic performance [9,10]. Moreover, these constructs play major roles in students' career decision-making process [11–14]. This holds true across global cultures, affirming that self-concepts and personal values matter more in determining whether students foresee themselves in science careers than do performance outcomes [15]. Considering the general dearth of individuals from statistically underrepresented communities pursuing physical science degrees in comparison to other STEM majors, more attention should be paid to these often-overlooked factors [16,17]. While focusing on the impact of active learning curricula on academic performance provides valuable support in favor of these methods, their possible effect on affect may help researchers better understand why students persist (or do not persist) in a major. In this paper we make the argument that an examination of self-efficacy formation as a factor of students' classroom interactions merits attention.

## II. THE NATURE OF SELF-EFFICACY

Of the constructs related to both performance attainment and career choice, self-efficacy plays a unique, well-tested, and strongly influential role [9,12–14,18–22]. Even while controlling for prior academic attainment, aptitude, and career interest, self-efficacy continues to significantly predict career choices [23,24]. When comparing self-efficacy to career-choice theories that link students to a vocation based on their personalities, self-efficacy has greater predictive power for career-choice than personality-based theories [25]. Students with high self-efficacy regarding tasks related to a field of study will more likely develop interests in, set goals toward, and make positive decisions about careers in that field [13]; indeed, many other factors play a similar role, but few hold the influence that self-efficacy does [19,26].

Bandura [27] describes self-efficacy as the beliefs individuals have about their capability to complete specific tasks and the outcomes they believe may result from their efforts. In other words, when faced with a particular problem to solve or assignment to accomplish, individuals make self-assessments about how successful they will be at solving said problem or completing said assignment. While somewhat akin to confidence and other expectancy constructs, self-efficacy differs in that the construct fluctuates according to both task and context [28]. For example, an individual may have one set of efficacy beliefs about her ability to solve a math problem, while holding a different set of beliefs about her ability to operate a voltmeter. Furthermore, the same individual may exhibit different self-efficacy beliefs about the same task according to her context (e.g., performing to an audience versus performing in private) [29].

Bandura proposes four types of experiences (i.e., sources) contribute to a person's self-efficacy beliefs: mastery experiences, vicarious learning, verbal persuasion, and physiological states [30]. Students' self-efficacy on physics related tasks is influenced by (i) students' past performance on similar tasks (i.e., mastery experiences), (ii) observations of peers to whom they relate succeeding or failing at those tasks (i.e., vicarious learning), (iii) direct encouragement or discouragement from peers, instructors, and others (i.e., verbal or social persuasion), and (iv) the emotional and physiological states of each student at the moment one assesses their self-efficacy or when students think about completing the task in question (i.e., physiological states). Anxiety, depression, or excitations are examples of physiological states that can contribute to students' self-efficacy.

### A. The social nature of self-efficacy development

Although individuals regulate their self-efficacy internally [10], some of the experiences that contribute to self-efficacy development result from external interactions in social settings. (Here and throughout the rest of the paper we define social settings as locations where two or more individuals work in close proximity on related tasks). We argue that development of efficacy beliefs, to an extent, relies on social interactions in these types of settings, which are the hallmark of various reformed physics courses. Our basis for this begins with how theory defines vicarious learning and verbal persuasion—two of the four established sources of self-efficacy. Vicarious learning and verbal persuasion experiences imply social settings.

Vicarious learning (VL) requires that an individual in question observes another person succeeding or failing at a given task. For this to occur, two or more persons must find themselves in the same space, within reasonable distance to observe one another's performance. While one may argue that this need not occur in physical proximity (e.g., watching videos of someone performing the task), the bulk of formal education environments primarily allow for in-person vicarious learning experiences.[1]

The presence of peers does more than create VL opportunities, it also nurtures threatening or affirming contexts that result in changes to students' overall self-efficacy. This holds particular sway in circumstances where individuals rate their performance by comparing their progress to that of those around them. In the case where a person observes others surpassing his or her performance, that individual has a higher likelihood of feeling less confident about his or her ability to perform the task at hand [29]. Educational settings often place students in situations where they find themselves explicitly or implicitly ranked among their peers according to their academic success. This ranking need not occur publicly or blatantly, but may be perceived by students nevertheless (e.g., a teacher drawing smiley faces on just a subset of graded exams).

Social interactions are also required in circumstances where individuals receive verbal feedback on performance, which may strengthen or undermine their self-efficacy. In general, classroom structures provide a forum for these kinds of verbal persuasion (VP) experiences to take place. Students often receive verbal recognition about their progress from teachers, peers, and on occasion, administrators. On a similar note, the type of emphasis placed on these performance evaluations matters [31]. Feedback that accentuates shortcomings contributes more to the breakdown of efficacy beliefs than feedback that focuses on amount of progress [32].

Some studies reveal that the socially oriented sources of self-efficacy (i.e., VL and VP) play a more significant role in the development and sustaining of women's efficacy beliefs [21]. For example, Zeldin and Pajares [20] interviewed 15 women working in STEM fields where underrepresentation of women persists, which included engineering and computer science. The researchers asked them about their self-beliefs and career history, specifically probing for information about their mathematics efficacy beliefs because of the highly mathematical nature of their professions. The participants reported experiences in line with VL and VP as playing a critical role in their career decision-making process and their persistence in their respective fields. These VL and VP experiences often included the presence of role models and positive encouragement from grade school teachers.

---

[1]The context of online education may limit such experiences and render this statement less valid.

### B. The social nature of Modeling Instruction

Of the existing, reformed instructional approaches directed at introductory university physics curricula (with Calculus) our research focuses on Modeling Instruction (MI), which differs significantly from the more common, lecture-based introductory-course format. MI introductory physics (referred to as "MI" from here on out) courses have tended to support low student-instructor ratios, short or nonexistent lectures, high numbers of solicited student-student and student-instructor interactions, and classroom settings designed to promote small group formation and collaborative learning. Students explore physical phenomena and solve classroom assignments in small groups, use various representations to summarize their conclusions on a white board, and come together during a "Board Meeting" to share and evaluate group solutions. Board Meetings—a characteristic feature of MI—reflect the highly social nature of learning that takes place [6].

The originators of MI developed this approach in order to promote student engagement for the purpose of mediating the construction of physics knowledge [33]. This grounding highlights the dialectical process where individuals reconcile their naïve ideas with concepts presented in the curriculum, which in this case occurs via experimentation and argumentation—the latter better described as social exchanges of ideas. Further development of MI by Desbien [34], as well as Brewe [6], cemented the inherently social nature of knowledge construction espoused by this physics teaching method. Grouping students, encouraging them to develop physics models together, and then having them relate group results to a larger classroom setting provides participants with opportunities to create knowledge and shared meanings or interpretations via verbal exchanges. This relationship between the building of knowledge and discussion is summarized in a common motto of the MI process: learning and social interactions are not mutually exclusive [35]. Additionally, learning occurs within a physics context. Students in this active learning environment employ a variety of physics-relevant tools, including language, to develop representations of physics concepts. For a detailed description of MI, please see Brewe [6].

Studies have shown that MI has led to increased student understanding in physics and improved attitudes toward physics [36,37]. Results documented in Brewe et al. [36] showed that students in MI courses have a 6.73 times greater odds of success than their counterparts in lecture sections. In addition to successfully passing, students in MI courses have greater pre-post gains on the Force Concept Inventory than students in traditional, lecture-based courses. The researchers observed these learning advantages for both women and men, though they note that the presence of a "gender gap" remains. Moreover, MI courses, unlike other successful, reformed physics approaches, positively shift student attitudes toward physics even when examined across varied instructors (Avg. effect size: Cohen's $d = 0.45$)—a feat accomplished by no other study known to us [37].

### C. Student self-efficacy in college physics

Research studies have reported correlations between self-efficacy and final grade in introductory physics courses, as well as the likelihood of passing the class [36,38,39]. The same can be said about other introductory courses in STEM fields, including chemistry, biology, and computer science [40–42]. Not only does physics self-efficacy impact academic performance, but it has also been shown to have a direct correlation with student affect, like motivation in physics courses [38].

Gender trends have also been reported, some of which are seen not only on self-efficacy as a whole, but also on the sources of self-efficacy. A study of 281 first-year college students that belonged to the same physics cohort revealed that female students reported lower self-efficacy beliefs than their male counterparts [43]. Moreover, in this same study, male students that had not taken any high school physics courses had higher self-efficacy than all other groups of students, indicating a gendered overconfidence. Larose et al. [41] performed a longitudinal study where female students who experienced increases in their self-efficacy during and after high school were more likely to report stability in their STEM-related vocational choices. This applied when controlling for high school achievement and socioeconomic status. On the other hand, a decline in self-efficacy had the opposite effect for female students and no effect on male students. In general, men who successfully attain STEM careers where underrepresentation of women exists report mastery experiences as the basis for their persistence and ongoing achievement [18]. Women in similar contexts, as noted earlier, seem to rely on VL and VP experiences as attributes of their professional success [20].

Previous studies on MI have explicitly explored students' self-efficacy [44,45]. A study by Sawtelle et al. [46] revealed that respondents taking one of several 30-student capacity MI courses at a public research university, regardless of gender, did not exhibit a statistically significant change in overall self-efficacy. When disaggregated by the sources of self-efficacy, the results did reveal an *increase* for women on the VP subscale. On the other hand, the same study revealed that both male and female students in lecture-based introductory physics courses exhibited a *drop* in self-efficacy. This drop held true across all four sources of self-efficacy. These findings align somewhat with findings by Fencl and Scheel [47] who showed that calculus-based physics I courses that employ a mixture of reformed pedagogical approaches, in particular student collaborations, have a stronger positive impact on students' self-efficacy than traditionally taught courses. This effect is enhanced for physics majors. Another study by Sawtelle *et al.* [44] employed logistic regression analysis to show that mastery experiences predict the rate at which male students pass or fail introductory physics, while female students' success depends more on vicarious learning.

Although the studies done with students participating in MI take a first step toward our understanding of self-efficacy development in these kinds of active learning environments, missing from the analyses are careful controls for other variables associated with self-efficacy, such as student ethnicity, as well as a more focused approach to understanding the role played by the MI curriculum's most prominent feature: social interactions. Considering additional limitations, such as potential selection bias introduced by the use of online surveys and the amount of unincorporated missing data, the propositions of the referenced studies in MI warrant further exploration. Moreover, our investigation will allow us to examine the effect of the curriculum in larger class-size settings.

## III. PURPOSE

This study aims to more carefully examine both changes in students' self-efficacy in a larger MI course as well as test our belief that the prevalent social interactions that occur in these courses have a notable relationship with self-efficacy development. This approach does not endeavor to compare MI to lecture-based pedagogies, but rather offers a more introspective look at the affective outcomes of MI as an active-learning curriculum. Using self-efficacy theory as a guide, we suggest that individual students come into class with certain internal expectations about their performance in the MI course. These expectations may differ according to each source of self-efficacy. For example, a student may have high expectation to receive praise from others (i.e., VP) but lower expectations to learn from peers (i.e., VL). Classroom experiences will influence students' expectations along the four sources of self-efficacy (see Fig. 1). We pay particular attention to VP and VL because the social nature of the MI curriculum leads us to hypothesize heightened prevalence for these events. We expect these types of experiences influence overall student self-efficacy at the end of the semester.

Given the increases of student conceptual understanding in MI courses, the social nature of self-efficacy development, and the highly interactive structure of MI courses, we

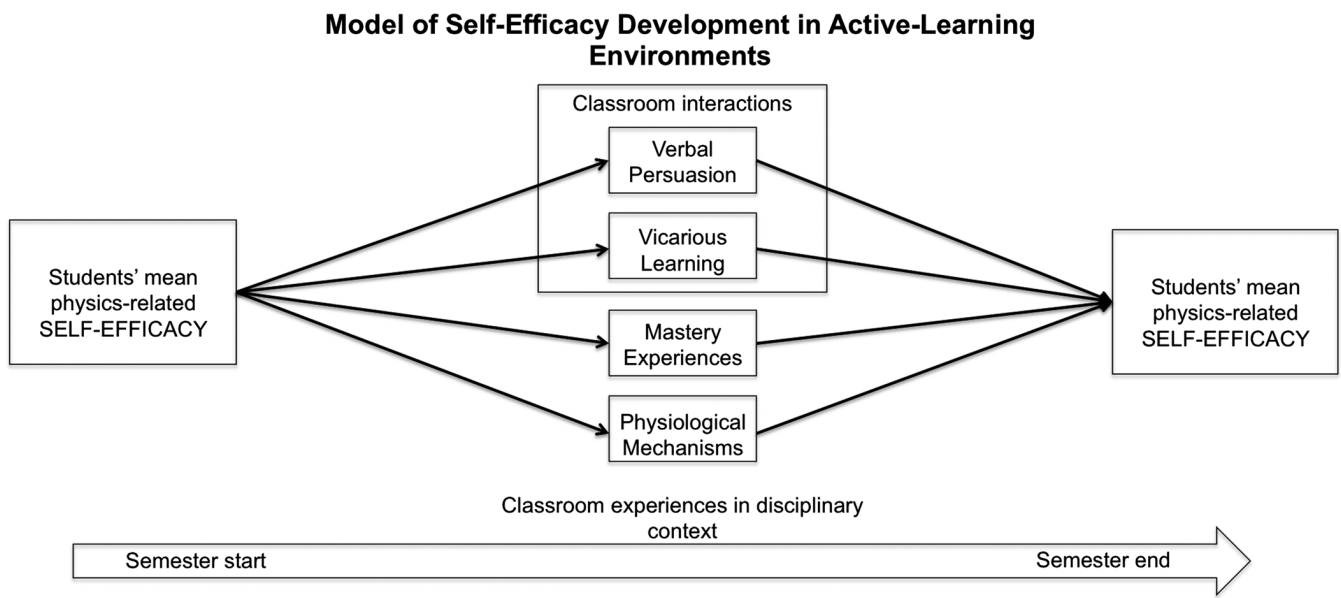

**Model of Self-Efficacy Development in Active-Learning Environments**

FIG. 1. Our model of self-efficacy development in active learning environments accounts for students' initial self-efficacy and its subsequent development as a result of classroom experiences. In alignment with theory, some of the development arises from learning experiences not directly related to social interaction (i.e., mastery experiences) [27]. In addition, we postulate that the social nature of many active learning environments has the capability of generating opportunities for students to receive verbal feedback or perceive others with whom they relate as successful or unsuccessful on physics tasks (i.e., verbal persuasion and vicarious learning experiences). Thus, we posit a link between certain types of classroom interactions and self-efficacy development.

hypothesized that students would exhibit a positive shift in their efficacy beliefs related to physics and the MI classroom even when controlling for variables associated with self-efficacy development. Furthermore, we use students' in-class social networks as a proximal measure of types and abundance of potential VL- and VP-related experiences that may play a role in mediating self-efficacy shifts. Specifically, we sought to address the following research questions:

1. Do students in the MI course experience statistically significant changes in physics self-efficacy as measured by PRE and POST scores on a self-efficacy in physics instrument (i.e., Sources of Self-Efficacy in Science Courses—Physics)?

2. Do students in the MI course experience statistically significant changes in physics self-efficacy scores when disaggregated by the four sources of self-efficacy?

3. How are social interactions as measured by student network centrality in the MI classroom associated to changes in students' self-efficacy?

4. Do other variables historically associated with student success in physics, such as gender, major, and ethnicity, contribute to the variance in students' POST self-efficacy scores when controlling for PRE scores?

## IV. A NOTE ON SOCIAL NETWORK ANALYSIS

Although the employment of social network analysis (SNA) in sociology has been taking place since the 1930s [48], its use in education research has experienced a growing popularity in recent years [35,49–52]. Discipline-based education researchers have explicitly encouraged the use of SNA to understand the social networks formed during learning [51]. Even more specific to the field of physics education research (PER), Bruun and Brewe [35] have suggested that increased application of SNA will better help the field understand student cognition. Many of the above cited papers may serve as primers to education researchers desiring to further their comprehension of SNA terminology and implementation. Grunspan *et al.* present a concise introduction targeted at science education researchers [51].

In brief, social network analysts endeavor to quantify the role of particular individuals in a network and the characteristics of a network and its evolution [53]. Our study focuses on measuring the "centrality" of actors in our network. Centrality can be calculated from students' interactions in a variety of ways. For example, the most basic form of centrality is "degree" centrality, which simply refers to the number of people with whom a person in a network interacts [35]. Other measures in the centrality family include inDegree, outDegree, PageRank, Closeness, and Betweenness. These may be calculated using the same student interaction information.

We collected student network data in order to calculate three specific measures of directed centrality: inDegree, outDegree, and PageRank. InDegree centrality measures direct incoming interactions (i.e., the number of times student $Y$ is listed by peers) and outDegree measures direct outgoing interactions and in some cases can be thought of as a measure of one's sociability (i.e., the number of peers student $Y$ lists). PageRank captures direct incoming interactions while taking into account the social connectedness of nodes leading to a student. PageRank offers a measure of weight to being named directly by a student who is often named by others. The PageRank algorithm establishes a node's importance using the number of links to the node, but also each node can then redistribute that importance by its number of outgoing links [35]. It is worth noting that students reported more often by others will have higher inDegrees and tend to have higher PageRanks. That is to say that inDegree and PageRank may be interpreted as measures of popularity or recognition from other actors in a social network since the more often a person is named, the more his or her PageRank grows. We chose to examine these three measures of centrality (i.e., inDegree, outDegree, PageRank) primarily because they limit our analysis of the relationship between self-efficacy and social interactions to students who had direct interactions with one another. They also follow with the uses and recommendations of past research [49–52], and they are generally understood by researchers outside the field of SNA. Moreover, by examining whether each of these three centrality measures contributes to changes in students' self-efficacy, we may get a clearer picture of the *kinds* of interactions that matter for student self-efficacy formation in MI courses.

Specifically, we secured responses from students on this question: "Name the individual(s) you had a meaningful classroom interaction with today." (see Sec. VI for more details about the network survey). Responses to this question can be used to calculate a plurality of network measures, not just the ones addressed in this study. Given student responses, InDegree can then be characterized as the number of incoming connections for a student. The number of participants a student reports or initiates interactions with is that person's outDegree. PageRank takes a more sophisticated approach to measuring the "importance" of a student or actor in a network. Developed by Brin and Page [54] for the Google search engine algorithm, the measure has been compared to calculating the probability of a random walker on a directed network to arrive at a particular node [55]. This means that not only does a node's inDegree affect its PageRank, but so does the inDegree of its neighbors (see Sec. VII. B.).

## V. CONTEXT

### A. Florida International University

Florida International University (FIU), Miami's public, urban research university, boasts a unique population. The institution educates over 56 000 students, making it one of the largest public universities in the country. Over 60% of FIU's students identify themselves as "Hispanic," while 13% identify themselves as "Black," another 12% as "White," and 13% as "Other" [56]. FIU is classified as a Hispanic Serving Institution (HSI), offering critically important services to the members of its community, which are primarily Hispanic. Considering recent national calls for a greater number of STEM majors, many of which include an emphasis on recruiting from underrepresented groups [8,16], it is relevant that no other university awards more STEM bachelor's degrees to underrepresented minorities than FIU [57].

## B. Introductory Physics I with Calculus at FIU

Students regardless of major or academic year have the option of self-selecting into one of the MI sections offered each semester or the lecture-based sections of Introductory Physics I with Calculus. The MI course incorporates the lab credit. It is worth noting that student familiarity with the MI approach varies. For example, students registered in the Fall 2015 MI courses responded differently to being asked about their expectations for the course. Of the 44 survey respondents, 9% expected a course only slightly different than lecture, 32% expected a much more interactive and hands-on experience, while a similar number expected no differences from a traditional lecture-based course. Remaining students either had no expectations or did not respond. Lecture sections at FIU usually have enrollments that range from 120 to nearly 400 students, though some offer a much lower class size limit. Students in a lecture section usually register concurrently for a respective laboratory course, but are not required to do so. In the Fall of 2014 only one section of MI was offered, limiting students' scheduling flexibility, but this particular section was the first designed to serve 75 students—over twice the number of students previously attempted—in a technology-saturated classroom specifically designed for active learning. Prior iterations of the course limited enrollment capacity at 30. Two sections of the large-capacity MI course were offered during the Fall 2015 term—one taught by the same experienced instructor who taught the Fall 2014 course and another taught by a postdoc. In order to accommodate the larger number of students, two graduate teaching assistants and three experienced learning assistants (i.e., undergraduate students) helped to facilitate instruction during courses in both terms. Only data from classes taught by the same primary instructor were used in this study in order to minimize confounding variables introduced by having data from different instructors.

## VI. METHODS

We obtained student data from FIU's database, which keeps a record of student responses to demographic questions answered at the time they apply to the university. Some of the majors represented in the courses included Engineering, Chemistry, Pre-Med, and English. No student in either MI course (i.e., Fall 2014 and Fall 2015) had declared physics as a major at the beginning of the semester, though we should note that students who declared dual majors were categorized under a larger umbrella (i.e., DUALFIU), which may include physics majors. The classes were composed of four prominent ethnic groups into which students identified: Asian, White, Hispanic, and Black. The majority of students enrolled in both classes identified themselves as Hispanic (47 women and 58 men), while eight identified themselves as Asian (three women and five men), 13 as White (five women and eight men), and 11 students as Black (two women and nine men). Four students identified as other or more than one race. The race and gender of the remaining six students in our data set were not available.

Self-efficacy surveys were administered in class on the first day of each semester (i.e., pre) and once during the last week of the semester (i.e., post). We had an overall 92% response rate on the pre based on a total of 147 students who registered for Fall 2014 and Fall 2015 MI courses. Our post administrations yielded an 80% response rate.

### A. Self-efficacy survey: Strengths and limitations

We employed the 33-item SOSESC-P survey to gauge the sources of self-efficacy and to get a measure of overall student self-efficacy. We chose this survey for a variety of reasons, including its specific designation for physics classroom settings given that self-efficacy measures require task-relevant items in order to align with the construct's definition [28]. The SOSESC-P was designed so that responses to statements can be disaggregated by each of the four sources of self-efficacy. We achieved an overall reliability alpha coefficient of 0.94, and reliability coefficients of 0.73 for verbal persuasion (7 items), 0.76 for vicarious learning (7 items), 0.84 for physiological mechanisms (9 items), and 0.86 for mastery experiences (10 items) subscales. These values align with past research led by the instrument's developers [39]. In that same study the survey was shown to correlate well with the Self-Efficacy for Academic Milestones Strength scale—a positively recognized and validated instrument. Some of the statements on the survey included the following: "*I am capable of receiving good grades on assignments in this class*" (mastery experience) and "*I will get positive feedback about my ability to recall physics ideas*" (verbal persuasion). Students used a five-point Likert scale to express agreement or disagreement with these. Overall scores in our study ranged from 79 to 165. The use of the SOSESC-P also supported continuity with past studies performed at FIU that employed the same instrument.

Though the SOSESC-P was designed for the purpose of measuring overall self-efficacy and the sources of self-efficacy, prominent researchers in the field warn about potential issues caused by combining two or more sources of self-efficacy [58]. These argue that combining items specific to each source increases ambiguity about what exactly is being measured and that students' context, including gender and ethnicity, may shift the combination of sources that contribute to students' actual self-efficacy. We present this as a limitation of our study and for that reason we report on analyses of each source of self-efficacy separately, in addition to students' total score on the SOSESC-P, which we interpret as a proxy for student self-efficacy. We do so on the grounds that we found significant change on all four sources of self-efficacy and criteria established by past studies [39,44,46,47,59].

### B. Social network survey

Since we could not directly measure when a student happens to have a meaningful VP or VL experience, we adopted an indirect approach that quantifies the number and types of social interactions students have using SNA. We also did this to test the model that the quantity and quality of certain kinds of interactions correlates with changes in students' self-efficacy and sources of self-efficacy. To measure relevant social interactions we administered a social network survey on the last day of the first week of class and subsequently once a month until the end of the semester for a total of 5 administrations. The development of this short survey took place under the guidance of the PER group at FIU, building off a previously used survey [49]. Of the open-ended questions appearing on this survey, only the first is relevant to this study: "Name the individual(s) (first and last name) you had a meaningful classroom interaction with today, even if you were not the main person speaking or contributing. (*You may include names of students outside of the group you usually work with*)." We provided a note to participants stating, "classroom interaction includes but is not limited to people you worked with to solve physics problems and people that you watched or listened to while solving physics problems." Blank space was provided so that participants could list as few or as many individuals they wished to. We carefully analyzed responses in order to identify the students listed. When 100% certainty or agreement could not be established as to the identity of a written name, a unique code was created for that specific report. This occurred five times when students with common first names were reported sans last name. To avoid this issue in the Fall 2015 course, we attached a numbered roster of students to the survey.

## VII. RESULTS

### A. Diagnosing changes in self-efficacy

Prior to performing $t$ tests we imputed student responses to the SOSESC-P in order to preserve the structure of our data, which reduces the rate of type I error by better accounting for nonresponses than would simply removing those cases from the analysis [60]. Multiple imputation is a Monte Carlo technique that replaces missing values using a likelihood function that assumes missing data is missing at random (MAR) and not because of reporting bias not captured by other variables [61]. For that reason we included responses to pre and post SOSESC-P surveys, student GPA at the start of the course, gender, and centrality measures when estimating values for the missing data. Given that we had no more than a 20% nonresponse rate on the SOSESC-P we ran five imputations ($m = 5$) as suggested by the literature using the *Amelia II* package [62] in R [63]. We ran the same analyses on all five data sets and pooled the results according to Rubin [64,65]. Since imputed values were generated for missing cases, the resulting $N$ (i.e., $N = 147$) included all unique participants enrolled in the fall courses during the first week of the semester.

We performed a dependent samples $t$ test to compare the mean total scores of the pre SOSESC-P responses ($M_{pre} = 135.36$, SD $= 13.86$) to those of the post ($M_{post} = 129.11$, SD $= 17.23$). The outcome revealed a statistically significant drop in physics-related self-efficacy from the beginning of the semester to the end of the semester [$t(146) = -4.75$, $p < 0.001$] with a small to medium effect size (Cohen's $d = 0.40$). In order to further explore the breakdown of students' sources of self-efficacy, we disaggregated responses on the SOSESC-P according to the following sources of self-efficacy: mastery experiences (ME), VL (i.e., vicarious learning), VP (i.e., verbal persuasion), and physiological states (PS). Dependent sample $t$ tests on each of these subsections showed a statistically significant drop in students' sources of self-efficacy on every portion of the survey even when setting our threshold alpha at 0.0125 in order to apply a Bonferroni correction to diminish type I error (see Table I).

### B. Measuring social interactions

We combined students' responses to the social network survey across the first four administrations. We did this with the goal of preserving uniformity of data collection. We planned for five survey administrations with the requirement that they take place during a typical MI class in which student groups work together on collaborative activities. Student interactions were primarily student generated and participants worked on physics related tasks. We achieved this setting across the first four data collections from both semesters in question, which had response rates of over 75%. Final exam scheduling altered the intended environment for the fifth administration both in the fall of 2014 and in the fall of 2015. Still, we pursued collection of data from the last survey, which was given during optional final exam review classes where students who chose to attend were not encouraged to participate in active-learning physics related inquiry. This is relevant since social context affects students' interactions, and the student network may reflect this. The uncharacteristic environment of this setting yielded less than 50% response rates and altered the resulting student network. In order to maintain fidelity of implementation, data from these surveys were not admitted into the final results, though analysis revealed nearly identical outcomes when included.

TABLE I. Although students in MI courses typically show conceptual and attitudinal gains, these results suggest that students in MI experience a statistically significant drop in physics self-efficacy. This drop also shows up significantly on all subsections of the SOSESC-P.

| Changes in sources of self-efficacy scores: | | | | |
|---|---|---|---|---|
| Dependent samples comparisons of SOSESC-P shifts (post-pre; $N = 147$) | | | | |
|  | Total score | ME | VL | PS | VP |
| Pre | 135.36 | 40.87 | 29.74 | 34.76 | 30.11 |
| *SD* | 13.86 | 4.71 | 3.30 | 4.82 | 2.85 |
| Post | 129.11 | 38.54 | 28.1 | 33.28 | 28.9 |
| *SD* | 17.23 | 6.05 | 4.05 | 6.44 | 3.51 |
| Diff. in mean | −6.25[a] | −2.33[a] | −1.64[a] | −1.48[a] | −1.21[a] |
| *t* value | −4.75 | −5.21 | −4.68 | −3.28 | −3.81 |
| Cohen's *d* | 0.40 | 0.43 | 0.39 | 0.27 | 0.32 |

[a]$p < 0.001$

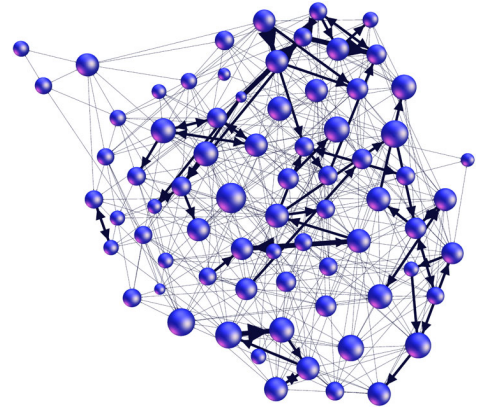

FIG. 2. Combined student network in the MI course for Fall 2014 drawn using the Force Atlas algorithm on Gephi [66]. Sphere size represents PageRank centrality and edge thickness represents weight of tie. Instructors have been removed.

From the responses to the network survey question we constructed directed edge lists indicating the source of the interaction (i.e., student responding to the survey) and each target listed on the survey (i.e., student name written in response to the question). The edge lists from the first four collections were combined and every interaction given a value of "1." Repeated interactions with the same targets were given a weight of $+1$ for each additional time the targets were listed on other administrations of the same survey question (see Fig. 2 for an example of the Fall 2014 network structure). After combining data from both semesters, students' total inDegree ($M = 14.1$, $\text{SD} = 6.04$), outDegree ($M = 18.2$, $\text{SD} = 10.4$), and directed PageRank ($M = 1.19 \times 10^{-2}$, $\text{SD} = 2.56 \times 10^{-3}$) were calculated in R using the *igraph* package [59]. InDegree was calculated by adding up the number of times a student was listed on question one of the four network surveys. OutDegree was calculated by adding up the number of individuals each particular student listed on question one of all four surveys, including instructors. Directed PageRank was calculated in *igraph* from incoming and outgoing links using the algorithm developed by Brin and Page [54] and represented by

$$p(i) = \frac{q}{n} + (1 - q) \sum_{j:\,j \to i} \frac{p(j)}{k_{\text{out}}(j)} \qquad i = 1, 2, \ldots, n, \quad (1)$$

where $p$ is the PageRank of node $i$, $j$ represents a node in the network linked to $i$, $p(j)$ and $k_{\text{out}}(j)$ are the PageRank and outDegree of node $j$, respectively, and $q$ is a damping factor commonly set at 0.15 as precedent in the literature [55,67].

We tested four linear regression models that aimed to predict total post self-efficacy scores while controlling for pre scores. Because network data often fails to meet the assumption of independence, measures of centrality often result in non-normal distributions. Bootstrapped linear regressions do not require assumptions about the distribution; therefore, we used this technique in order to account for any dependency in data retrieved from the social network [68]. Bootstrapping is a Monte Carlo approach that applies a random resampling of the existing data set to calculate a set of regression coefficients on that sample. We did so over 1000 iterations on each dataset and created a distribution of coefficients by which to compare the values in our data [69]. 95% confidence intervals (CI) for our parameters were calculated using the bias-corrected and accelerated method developed by Efron [70], which better addresses bias and skewness while producing narrower intervals. These analyses were run on each of our imputations with nearly identical results, which were then pooled. Although, in general, all four models predicted the dependent variable, the models revealed that PageRank was the only statistically significant predictor besides the control variable. Regression coefficients for inDegree and outDegree had confidence intervals that included zero. PageRank explained an additional 3.7% of the variance in students' post self-efficacy scores (see Table II). Because of potential collinearity between the centrality measures, we tested these variables using separate models. The correlation between PageRank and inDegree was 0.46 ($p < 0.001$), between PageRank and outDegree was 0.24 ($p < 0.01$), and between inDegree and outDegree was 0.76 ($p < 0.001$). Again, because centrality measures typically fail to meet the assumption of normality required by traditional statistical tests, the above correlations were calculated using a permutation test for correlation, which also employs a Monte Carlo method.

TABLE II.   Models using network variables predicting post self-efficacy scores. InDegree and PageRank centralities capture a measure of recognition, but PageRank weighs that recognition according to the popularity of peers interacting with the student. Here we show that only PageRank predicts overall self-efficacy scores. Note standardized regression coefficients (i.e., $\beta$) appear in parentheses.

| Model-level statistics | | | |
|---|---|---|---|
| $F$ statistic | $F(1, 111) = 42.34$ | $F(2, 110) = 25.04$ | $F(2, 110) = 22.58$ | $F(2, 110) = 22.76$ |
| R square | 0.276 | 0.313 | 0.291 | 0.293 |
| 95% CI for R square | (0.114, 0.445) | (0.159, 0.472) | (0.122, 0.453) | (0.128, 0.451) |
| Regression coefficients | | | |
| Predictors | Model 1 | Model 2 | Model 3 | Model 4 |
| Pre SOSESC-P | 0.65 ($\beta_1 = 0.52$) CI[0.41, 0.88]; SE = 0.12 | 0.65 ($\beta_1 = 0.52$) CI[0.42, 0.88]; SE = 0.12 | 0.64 ($\beta_1 = 0.51$) CI[0.40, 0.87]; SE = 0.12 | 0.63 ($\beta_1 = 0.51$) CI[0.38, 0.85]; SE = 0.12 |
| PageRank | | 1380 ($\beta_2 = 0.21$) CI[279, 2539]; SE = 586 | | |
| inDegree | | | 0.37 ($\beta_2 = 0.13$) CI[−0.13, 0.80]; SE = 0.24 | |
| outDegree | | | | 0.22 ($\beta_2 = 0.13$) CI[−0.05, 0.47]; SE = 0.13 |

Additional models were tested to determine whether PageRank, inDegree, or outDegree centrality contributed to variance on the sources of self-efficacy. Each model tested whether a centrality measure predicted outcomes on the disaggregated postscore from each of the subsections of the SOSESC-P while controlling for the prescores for each subsection. Each measure of centrality was tested separately for each of the four subsections. Bootstrapped linear regressions indicated that although all models resulted in significant $F$ statistics with $p$ values of less than 0.001, only PageRank predicted postscores on the ME subsection ($B = 421$, $\beta = 0.17$, CI[44, 835], SE = 197), inDegree predicted some of the variance in postscores on the VP subsection ($B = 0.11$, $\beta = 0.17$, CI[0.01, 0.22], SE = 0.05), and outDegree predicted some of the variance in postscores on the VP ($B = 0.06$, $\beta = 0.16$, CI[0.01, 0.12], SE = 0.03) and VL ($B = 0.06$, $\beta = 0.14$, CI[0.001, 0.13], SE = 0.03) subsections. No centrality measure predicted students' post PS scores in a statistically significant way when controlling for prescores. Prescores on each section always predicted postscores. These results lend credence to the argument that students' relational position in the social network of a classroom is associated with changes in the sources of self-efficacy even when including nonsocial sources of self-efficacy, like mastery experiences. We address this further in our Discussion section.

### C. Examining other relevant variables

In order to gauge whether changes in students' self-efficacy scores were related to the presence of other variables associated with student performance, we undertook several additional analyses. Two separate student's independent samples $t$ tests were run to determine whether or not a difference exists between female and male students' pre- and postscores on the SOSESC-P. The analysis revealed that no statistically significant gender difference existed at the start of the MI courses or at the end. The same held true when examining the disaggregated sources of self-efficacy. Furthermore, a multiple linear regression model was examined to determine the ability of ethnicity and major, along with gender, to predict the variance in student self-efficacy scores at the end of the course when controlling for prescores. The results showed that the model was statistically significant ($p < 0.001$), but the only variable that contributed to the model's significance was prescore. Neither ethnicity nor declared major contributed to the variance in students' post self-efficacy scores, though to be sure, the low number of representatives from certain ethnic groups (e.g., Black) and majors (e.g., English) limited the power of our model and our ability to make strong claims about the effect of ethnicity and major. Given that gender differences were not seen on pre- and post self-efficacy scores, we did not expect this variable to be significant.

Bootstrapped analyses revealed no difference between the mean outDegree nor PageRank of female and male students. Nevertheless, male students on average had slightly higher inDegrees than did female students [$t(121.9) = -2.13$, $p < 0.05$, Cohen's $d = 0.37$; see Table III].

### VIII. DISCUSSION

Our examination of an active-learning, introductory physics course format revealed that regardless of gender, major, and ethnicity, students had on average lower beliefs about their ability to successfully complete physics related tasks at the end of the semester than they did at the beginning. This negative change was seen across the self-efficacy survey as a whole and when disaggregated by the four accepted sources of self-efficacy. Students report a decrease in the kinds of experiences that theoretically contribute to positive self-efficacy formation. This contrasts with a previous study in smaller classrooms using the same MI curriculum that showed no change in overall student self-efficacy and an increase along the verbal persuasion scale [46]. We suggest as a possibility that these differences may exist for several reasons, including class size and data structure (e.g., handling of missing data). Moreover, we set our alpha levels at much steeper thresholds in order to combat type I error—a correction this prior study did not apply. However, the drop we found is relatively small compared to the range of the self-efficacy scale and the variance in student responses. The drop may simply reflect a correction of students' overconfidence [43].

TABLE III. Gender-based comparisons of network centrality. InDegree and PageRank centralities do not differ significantly by gender. On the other hand, female students report more peers (i.e., outDegree) in response to the network survey question examined.

|  | inDegree | outDegree | PageRank |
|---|---|---|---|
| Mean differences (female–male) | $-2.13^{\text{a}}$ | $-0.63$ | $0$ |
| $T$ statistic | $-2.14$ | $-0.49$ | $0.74$ |
| Cohen's $d$ | $0.37$ | $0.09$ | $0.13$ |

$^{\text{a}}p < 0.05$

In light of past research on student academic outcomes in MI, what captures our interest is that students experienced a decrease as opposed to an increase in all the sources of self-efficacy. In fact, we hypothesized increases both on self-efficacy as a whole and on each of the four sources. The decrease found was approximately 73% as large as decreases seen in past studies with students in lecture courses [46]. These contrary results point to the need for further exploration of this topic, in particular with regard to factors that mediate these shifts. We should also note that our students started at higher levels than previously reported studies using the SOSESC-P [46,47]. While a variety of variables may have contributed to this latter attribute, any justification would merely be speculative.

Given the inherently social aspects of self-efficacy development in addition to the emphasis on discourse-based learning in the MI curriculum, we tested whether students' social behavior predicted self-efficacy shifts. We aver that a relationship exists between at least one kind of interaction, as captured by student PageRank centrality, and changes in students' overall efficacy beliefs. We found that the number of times a student is listed by popular peers makes a difference (see PageRank in Table II). That is to say that being named by a student whom others report having a high number of interactions with positively predicts increases in overall self-efficacy. In short, a 1 standard deviation increase in student PageRank results in a 0.21 standard deviation increase in post self-efficacy after controlling for prescores (see Table II). On the other hand, we did not find that the number of peers a student has a meaningful interaction with (i.e., outDegree) nor the number of times a student is recognized by his or her peers as having contributed to a meaningful interaction (i.e., inDegree) affect changes on the self-efficacy scale as a whole.

With regard to the sources of self-efficacy, PageRank also positively predicted mastery experience scores. This deserves some unpacking, as this source of self-efficacy is not typically associated with social interactions, but often plays a primary role in self-efficacy formation, especially for men [18]. Moreover, the number of both incoming and outgoing interactions positively predicted verbal persuasion scores, while only outgoing interactions positively predicted vicarious learning scores. None of the interactions examined had a statistically significant association with students' physiological state.

These results align with our model of self-efficacy development in active learning environments (see Fig. 1), but also expand on it. They support our belief that specific kinds of social academic experiences, as quantified using centrality measures, partially predict students' postmeasures on the inherently social sources of self-efficacy. Yet, the analyses also support expansion of the model as centrality was found to have an even stronger relationship with ME, which we did not consider as a source of self-efficacy related to social networks. In other words a student exhibiting a drop because of having poor results on a mastery experience (e.g., exam) did not necessarily strike us as an experience *directly* related to the student's network of peers. Nevertheless, indirectly, it may be possible that access to a support group in the class may provide students with capital that leads to improved performance as implied by previous studies on teacher networks and capital theory [71,72].

Although our linear models only explain a relatively small portion of additional variance, they forge a valuable link between SNA and the sources of self-efficacy. As expected, an increase in the number of times peers interact with a particular student increases the chances this student has positive verbal persuasion experiences. The specific items on the SOSESC-P suggest that the student is receiving encouragement about his or her physics ability. This aligns with the fact that others are reporting having salient academic interactions with this student. The same occurs with regard to a student's outDegree, but this kind of outgoing interaction—in the sense that it represents how often students reach out to peers—is also positively related with vicarious learning experiences. Since vicarious learning experiences theoretically indicate situations where one learns from watching someone with whom one relates, it is possible that these individuals are the ones students seek out. Although the data in this particular study do not allow us to make a definitive conclusion in that regard, they certainly offer some value to an examination of the kinds of individuals different students reach out to. This is further supported by the observed relationship between PageRank and mastery experiences. PageRank does more than simply tally the number of social interactions (i.e., outgoing or incoming), but also captures *with whom* the interaction occurs. Interactions coming from popular individuals as defined by their inDegree positively predict a students' sense that they can learn and get good grades in physics. Because students did not know each other's inDegree, we can infer that students recognize, in some capacity, who these popular individuals may be and have a perception about their academic popularity. A highly social setting may catalyze these peer-to-peer judgments.

Active-learning environments, like MI, create the kind of social space that allows students the flexibility to interact in different ways with different people [45]. Though no relationship was found between gender and self-efficacy, female and male students differ in the kinds of interactions they experience. Male students in this class are the subjects of others' meaningful interactions more so than female students. While we did not intend to focus on gender differences, we do present these results as evidence that certain students experience the social aspects of this type of environment differently. In our case, major and ethnicity did not contribute to these differences, but that may have been a result of our relatively low sample size in certain subgroups. The value of having examined several measures of centrality is justified in our ability to conclude that the types of interactions students experience and with whom they have these interactions matters with regard to self-efficacy formation. The characteristic of PageRank as a measure of the kinds of people whom students interact with may also help to explain why PageRank is a slightly better predictor of overall self-efficacy than inDegree or outDegree. Additionally, we know from past studies that mastery experiences, a source we found associated with PageRank, often plays a greater role in self-efficacy formation in physics courses than other sources [18,44].

Our surprising results encourage us to think about ways to mitigate effects of the social structure of MI on students' efficacy beliefs and vice versa. This might manifest itself through the purposeful stimulation of interactions between certain groups of students. Altering how students participate in the social aspects of a classroom in a way that gives all an equitable chance then becomes, in part, an issue of how students recognize the value of their peers. We suspect that the highly social nature of this learning approach exposes students to academic judgment from peers and can initiate introspective evaluation, specifically while students solve problems in groups and when they present solutions to the larger classroom. The increased number of interaction events may provide students with more opportunities to generate perceptions about their peers' ability to contribute to a physics-related task and, in turn, influence whom they work with or whom they list when asked to recall meaningful academic interactions. These perceptions can drive changes in interactions. Although in this example we have suggested that these changes may relate to academic perceptions, they may also relate to students' ability to communicate effectively, helpfulness, or even friendliness.

We faced certain limitations worth noting. No student in the Fall 2014 and 2015 course had declared physics as a sole major. Physics majors may be less susceptible to changes in self-efficacy via peer recognition because of their strong physics identity relative to those pursuing other STEM fields [11]. The absence of physics majors in these MI courses might also point to a possibly unidentified source of self-selection bias. Furthermore, the MI classrooms in question were among the first at FIU to host that many students at once. The novelty of implementing this curriculum with more students in a brand new classroom may have led to unrecognized shortcomings. Further investigation should take place to more clearly understand how these factors relate to our study.

Knowing the powerful role that introductory physics courses play on career persistence and the underrepresentation of certain groups of students [7], we are pressed to search for ways to ensure that students complete the semester feeling more confident in their ability to perform physics tasks rather than less confident—regardless of the gradient. Though we report a somewhat minor 3.79% overall drop in students' physics self-efficacy, this is an average measure. Individually, students ranged from a 29% decrease from prescore to a 46% increase from prescore. This variance offers a living example of how students in the course can exhibit contrary, affective outcomes. Our study showed that part of what accounted for these differences are the kinds of interactions students had. Similar learning environments, particularly those that focus on active-learning mediated by student interactions, may exhibit parallel outcomes. Our holistic approach to student learning motivates us to explore ways to improve MI and interactive-learning approaches in the introductory classroom such that the maximal number of students leave not just academically prepared, but also affectively equipped to persist in physics careers. Our study aims to highlight the value of examining these facets of student outcomes in these environments, specifically self-efficacy development and course-related social interactions. It is not enough to simply say that students are learning more. This is especially true in the realm of career decision-making where self-efficacy plays a central role even for STEM related professions, partially explaining the underrepresentation of certain groups in these fields [73]. Bandura [29] explains,

*"…the stronger people's belief in their efficacy, the more career options they consider possible, the greater the interest they show in them, the better they prepare themselves educationally for different occupations, and the greater their staying power and success in difficult occupational pursuits."*

Our exploration of this matter reflects our commitment to not only help our students better understand physics, but also motivate some to join the physics community.

This requires that we focus on more than just content matter.

## ACKNOWLEDGMENTS

This work exists thanks in part to intellectual contributions from the Physics Education Group at Florida International University, as well as support from the FIU Division of Research and the STEM Transformation Institute at FIU. This study was made possible by National Science Foundation Grant No. PHY 1344247.